\documentclass[notitlepage,a4,10.5pt]{article}

\usepackage{graphicx}

\usepackage{amsmath}%
\usepackage{amsfonts}%
\usepackage{amssymb}%
\usepackage{mathrsfs}
\usepackage{amsthm}
\usepackage{float}

\usepackage{authblk}
\usepackage{enumitem}

\usepackage[left=2.5cm,right=2.5cm,top=3cm,bottom=3cm]{geometry} 

\usepackage{xcolor}
\usepackage{appendix}

\newcommand{\mc}{\mathcal}
\newcommand{\cp}{\times}

\newcommand{\bol}{\boldsymbol}

\newcommand{\abs}[1]{\left\lvert{#1}\right\rvert}

\newcommand{\lr}[1]{\left({#1}\right)}

\newcommand{\mf}{\mathfrak}
\newcommand{\p}{\partial}

\newcommand{\ti}[1]{\textit{#1}}
\newcommand{\tb}[1]{\textbf{#1}}

\begin{document}

\title{Existence of Weakly Quasisymmetric Magnetic Fields\\ in Asymmetric Toroidal Domains \\ with Non-Tangential Quasisymmetry}
\author[1]{Naoki Sato} 
\affil[1]{Graduate School of Frontier Sciences, \protect\\ The University of Tokyo, Kashiwa, Chiba 277-8561, Japan \protect\\ Email: sato\_naoki@edu.k.u-tokyo.ac.jp}
\date{\today}
\setcounter{Maxaffil}{0}
\renewcommand\Affilfont{\itshape\small}

    \maketitle
    \begin{abstract}
    A quasisymmetry is a special symmetry that enhances the ability of a magnetic field to trap charged particles.  
    Quasisymmetric magnetic fields 
    may allow the realization of next 
    generation fusion reactors (stellarators) with superior performance when compared with classical (tokamak) designs.  Nevertheless, the existence of such magnetic configurations lacks mathematical proof due to the complexity of the governing 
    equations. Here, we prove the existence of weakly quasisymmetric magnetic fields by constructing explicit examples. This result is achieved by a tailored parametrization of both magnetic field and hosting toroidal domain, which are optimized to fulfill quasisymmetry.   
    The obtained solutions hold in a toroidal volume, are smooth, 
    possess nested flux surfaces, are not invariant under continuous Euclidean isometries, have a non-vanishing current, exhibit a direction of quasisymmetry that is not tangential to the toroidal boundary, and fit within the framework of anisotropic magnetohydrodynamics. 
    \end{abstract}

\section{Introduction}

\begin{figure}[h]
\hspace*{-0cm}\centering
    \includegraphics[scale=0.4]{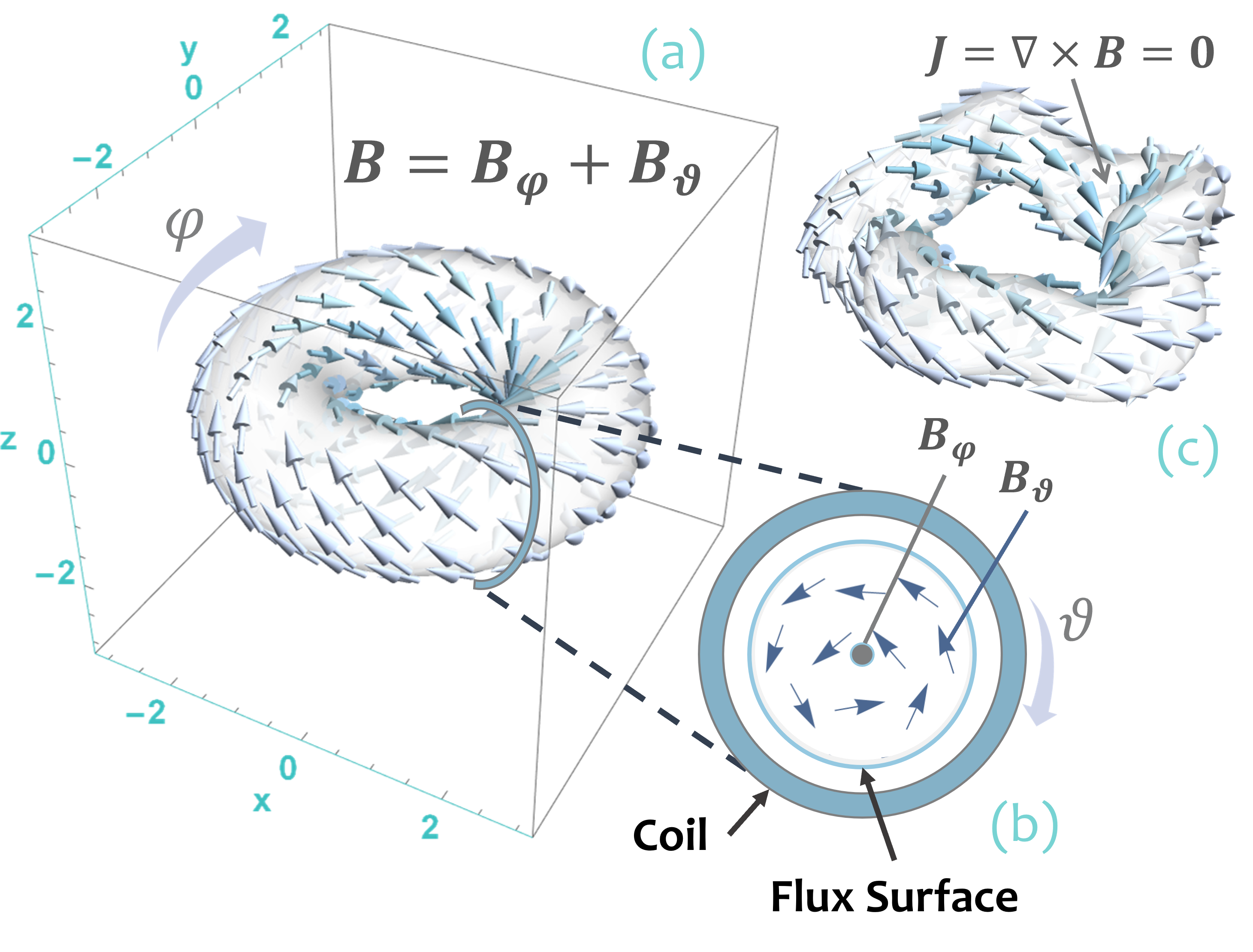}
\caption{\footnotesize (a) and (b): magnetic field configuration in an axially symmetric tokamak. The total confining magnetic field $\bol{B}=\bol{B}_{\varphi}+\bol{B}_{\vartheta}$ is given by an axial (toroidal) component $\bol{B}_{\varphi}$ produced by external coils plus a poloidal component $\bol{B}_{\vartheta}$ 
generated by an electric current flowing in the $\varphi$-direction. This current is sustained by the confined plasma itself. Here, $\varphi$ and $\vartheta$ denote toroidal angle and poloidal angle respectively. For simplicity, the reactor vessel separating external coils from the confinement region is not shown.  (a) The total magnetic field $\bol{B}$ over a flux surface $\Psi={\rm constant}$ such that $\bol{B}\cdot\nabla\Psi=0$. (b) 
Schematic view of toroidal component $\bol{B}_{\varphi}$ and poloidal component $\bol{B}_{\vartheta}$ on a cross section $\varphi={\rm constant}$. (c) Schematic representation of a stellarator: the confining magnetic field is asymmetric and entirely produced by external coils, implying that the associated electric current vanishes in the confinement region, $\bol{J}=\nabla\cp\bol{B}=\bol{0}$.}
\label{fig1}
\end{figure}

Nuclear fusion is a technology with the potential to revolutionize 
the way energy is harvested. 
In the approach to nuclear fusion based on magnetic confinement,
charged particles (the plasma fuel) are trapped in a doughnut-shaped (toroidal) reactor with the aid of a
suitably designed magnetic field. 
In a classical tokamak \cite{Wesson}, the reactor vessel is axially symmetric (see figure \ref{fig1}(a)). 
The axial symmetry is mathematically described 
by the independence of physical quantities, such as the magnetic field $\bol{B}$ and its modulus $B$, from the toroidal angle $\varphi$. 
Such symmetry is crucial to the quality of tokamak confinement,
because it ensures the conservation of the angular momentum $p_{\varphi}$ of charged particles. 
However, the constancy of $p_{\varphi}$ is not enough to constrain 
particle orbits in a limited volume because,  
in addition to the tendency to follow magnetic field lines, 
particles drift across the magnetic field. 
This perpendicular drift eventually causes particle loss
at the reactor wall, deteriorating the confinement needed to sustain fusion reactions. 
In a tokamak, perpendicular drifts are therefore suppressed by driving an axial electric current 
through the confinement region, which generates a poloidal magnetic field in addition 
to the external magnetic field produced by coils surrounding the confinement vessel (see figures \ref{fig1}(a) and \ref{fig1}(b)). 
The overall magnetic field therefore forms twisted helical field lines around the torus.  
Unfortunately, the control of such electric current is difficult because it is maintained 
by the circulation of the burning fuel itself, making steady operation of the machine a practical challenge. 

In contrast to tokamaks, stellarators \cite{Spitzer,Helander14} 
are designed to confine charged particles through a vacuum magnetic field 
produced by suitably crafted asymmetric coils (see figure \ref{fig1}(c)). 
In this context, symmetry is defined as invariance under 
continuous Euclidean isometries, i.e. transformations of three-dimensional
Euclidean space that preserve the Euclidean distance between points. 
In practice, these transformations are combinations of translations and rotations, 
with three 
corresponding 
types of symmetry: translational, rotational (including axial), and helical. 
The magnetic field generated by the asymmetric coils of a stellarator 
is endowed with the field line twist required to minimize 
particle loss associated with perpendicular drift motion. 
This removes,  
in principle, the need to drive an electric current within the confinement region,  
and thus enables the reactor to operate in a condition close to a steady state 
(in practice currents may exist in stellarators as well, but they are sensibly smaller than those in a tokamak).
Unfortunately, the loss of axial symmetry comes at a heavy price: 
in general, the angular momentum $p_{\varphi}$ is no longer constant, 
and confinement is degraded. 
However, a conserved momentum that spatially constrains 
particle orbits can be restored if the  magnetic field 
satisfies a more general kind of symmetry, the so-called quasisymmetry \cite{Helander14,Cary}.
The essential feature of a quasisymmetric magnetic field, whose rigorous definition \cite{Rodriguez2020} is given in equation \eqref{QS},
is the invariance $\bol{u}\cdot\nabla B=0$ of the modulus $B=\abs{\bol{B}}$ 
in a certain direction in space $\bol{u}$ (the quasisymmetry). 
For completeness, it should be noted that there exist two kinds of quasisymmetry \cite{Rodriguez2021QS,Burby2021,Burby2020,Tessa}: 
weak quasisymmetry (the one considered in the present paper), 
and strong quasisymmetry. In the former, quasisymmetry results in a conserved momentum 
at first order in the guiding-center expansion, 
while in the latter the conservation law originates from an exact symmetry
of the guiding-center Hamiltonian. 
Furthermore, the notion of quasisymmetry can be generalized to
omnigenity, a property that guarantees the suppression of perpendicular drifts on average \cite{Landre12}.

Despite the fact that several stellarators aiming at quasisymmetry or omnigenity have been built \cite{Canik,Pedersen}, 
that significant efforts are being devoted to stellarator optimization (see e.g. \cite{Bader}),  
and that quasisymmetric magnetic fields have been obtained with high numerical accuracy \cite{Landre22}, 
at present the existence of quasisymmetric magnetic fields lacks mathematical proof.
This deficiency is rooted in the complexity of the partial differential equations  
governing quasisymmetry, which are among the hardest in mathematical physics. Indeed, on one hand the toroidal volume where the solution is sought is itself a variable of the problem. On the other hand, the first order nature of the equations prevents general results from being established beyond the existence of local solutions. 
The availability of quasisymmetric magnetic fields also strongly depends on the
additional constraints that are imposed on the magnetic field. 
For example, if a quasisymmetric magnetic field is sought within the framework of ideal isotropic magnetohydrodynamics, 
the analysis of \cite{Garren} suggests that such configurations do not exist (see also \cite{Sengupta, Plunk, Constantin2021JPP, Constantin2021CMP}) due to an overdetermined system of equations where geometrical constraints outnumber the available degrees of freedom. 
The issue of overdetermination is less severe \cite{Rodriguez2021_1,Rodriguez2021,Rodriguezpre} if 
quasisymmetric mgnetic fields correspond to equilibria of
ideal anisotropic magnetohydrodynamics \cite{Grad66,Dobrott,Iacono} where scalar pressure is replaced by a pressure tensor. 
In this context, it has been shown \cite{Sato21} that local quasisymmetric magnetic fields do exist, 
although the local nature of the solutions is exemplified by a lack of periodicity around the torus.

The goal of the present paper is to establish the existence of weakly quasisymmetric magnetic fields
in toroidal domains by constructing explicit examples.  
This `constructive' approach has the advantage of bypassing the 
intrinsic difficulty of the general equations governing quasisymmetry, and hinges upon the method of Clebsch parametrization \cite{YosClebsch}, which provides an effective representation of the involved variables, including the shape of the boundary enclosing the confinement region. 
The quasisymmetric magnetic fields reported in the present paper hold within asymmetric toroidal volumes, are smooth, have nested flux surfaces, 
are not invariant under continuous Euclidean isometries, and can be regarded as quilibria of ideal anisotropic magnetohydrodynamics. 
Nevertheless, these results come with two caveats: since the constructed solutions are 
optimized only to fulfill weak quasisymmetry, the resulting magnetic fields are not vacuum fields, 
and their quasisymmetry does not lie on toroidal flux surfaces. 
Whether these two properties are consistent with weak quasisymmetry 
therefore remains an open theoretical issue.


\section{Construction of Quasisymmetric Magnetic Fields}
Let $\Omega\subset\mathbb{R}^3$ denote a smooth bounded domain with boundary $\p\Omega$.  
In the context of stellarator design $\Omega$ represents the volume occupied by the magnetically confined plasma,  
while the bounding surface $\p\Omega\simeq {\rm T}^2$ has the topology of a torus (a 2-dimensional manifold of genus 1).
It is important to observe that, in contrast with conventional tokamak design, the vessel $\p\Omega$ of a stellarator
does not exhibit neither axial nor helical symmetry.  
In $\Omega$, a stationary magnetic field $\bol{B}\lr{\bol{x}}$ is said to be weakly quasisymmetric provided that there exist a vector field $\bol{u}\lr{\bol{x}}$ and a function $\zeta\lr{\bol{x}}$ such that the following system of 
partial differential equations holds,
\begin{subequations}
\begin{align}
\nabla\cdot\bol{B}&=0,~~~~\bol{B}\cp\bol{u}=\nabla\zeta,~~~~\nabla\cdot\bol{u}=0,~~~~\bol{u}\cdot\nabla B^2=0~~~~{\rm in}~~\Omega,\label{QSa}\\
\bol{B}\cdot\bol{n}&=0~~~~{\rm on}~~\p\Omega,\label{QSb}
\end{align}\label{QS}
\end{subequations}
where $B=\abs{\bol{B}}$ is the modulus of $\bol{B}$, $\bol{n}$ denotes the unit outward normal to $\p\Omega$, and $\bol{u}$ is the direction of quasisymmetry. 
As previously explained, system \eqref{QSa} ensures the existence of a conserved momentum at first order in the
guiding center ordering that is expected to improve particle confinement. 
Usually, the function $\zeta$ is identified with the flux function $\Psi$ so that both $\bol{B}$ and $\bol{u}$ lie on flux surfaces $\Psi={\rm constant}$ and the conserved momentum originating from the quasisymmetry is well approximated by the flux function. Although this property is highly desirable from a confinement perspective because it confines particle orbits into a bounded region, 
if only weak quasisymmetry \eqref{QS} is sought $\zeta$ and $\Psi$ may differ (see e.g. \cite{Rodriguez2020}). 
In particular, allowing configurations with $\zeta\neq\Psi$ leaves the interesting possibility of achieving good confinement 
if the level sets of $\zeta$ enclose bounded regions with a topology that may depart from a torus. 
Mathematically, the four equations in system \eqref{QSa} represent so-called Lie-symmetries of the solution, i.e. 
the vanishing of the Lie-derivative $\mf{L}_{\bol{\xi}}T$ quantifying the infinitesimal difference between the value of a tensor  field $T$ at a given point and that obtained by advecting the tensor field along the flow generated by the vector field $\bol{\xi}$. 
Specifically, the first equation and the third equation, which imply that both $\bol{B}$ and $\bol{u}$ are
solenoidal vector fields, express conservation of volumes advected along $\bol{B}$ and $\bol{u}$ 
according to $\mf{L}_{\bol{B}}dV=\mf{L}_{\bol{u}}dV=\lr{\nabla\cdot\bol{B}}dV=\lr{\nabla\cdot\bol{u}}dV=0$, where $dV=dxdydz$ is the volume element in $\mathbb{R}^3$. Similarly, the second equation in \eqref{QSa} expresses the invariance of 
the vector field $\bol{B}$ along $\bol{u}$ according to $\mf{L}_{\bol{u}}\bol{B}=\bol{u}\cdot\nabla\bol{B}-\bol{B}\cdot\nabla\bol{u}=\nabla\cp\lr{\bol{B}\cp\bol{u}}=\bol{0}$, while the fourth equation 
expresses the invariance of the modulus $B^2$ along $\bol{u}$, i.e. $\mf{L}_{\bol{u}}B^2=\bol{u}\cdot\nabla B^2=0$.  
For further details on these points see \cite{Sato21}.

The construction of a solution of \eqref{QS} requires the simultaneous optimization of $\bol{B}$, $\bol{u}$, $\zeta$ and the shape of the boundary $\p\Omega$. 
Indeed, assigning the bounding surface $\p\Omega$ from the outset   
will generally prevent the existence of solutions due to overdetermination (the available degrees of freedom are not sufficient to satisfy the quasisymmetry equations).
A convenient way to simultaneously optimize $\bol{B}$, $\bol{u}$, $\zeta$, and $\p\Omega$ is to use Clebsch parameters \cite{YosClebsch}, which enable the enforcement of the topological requirement on $\p\Omega$, which must be a torus, and the extraction of the remaining geometrical degrees of freedom for  $\bol{B}$, $\bol{u}$, and $\zeta$. 
To see this, first observe that the unit outward normal $\bol{n}$ to the boundary $\p\Omega$ can be expressed through
the flux function $\Psi$, which is assumed to exist, according to $\bol{n}=\nabla\Psi/\abs{\nabla\Psi}$. Next, parametrize $\bol{B}$ and $\bol{u}$ as 
\begin{equation}
\bol{B}=\nabla \beta_1\cp \nabla\beta_2 ,~~~~\bol{u}=\nabla u_1\cp\nabla u_2,\label{Cl}
\end{equation}
where the Clebsch parameters $\beta_1$, $\beta_2$, $u_1$, and $u_2$ are (possibly multivalued) functions that must be determined from the quasisymmetry equations \eqref{QS}. 
Here, it should be noted that, due to the Lie-Darboux theorem \cite{DeLeon}, for a given smooth solenoidal vector field $\bol{v}$ one can always find single valued functions $\alpha_1$ and $\alpha_2$ defined in a sufficiently small neighborhood $U$ of a chosen  point $\bol{x}\in\Omega$ such that $\bol{v}=\nabla\alpha_1\cp\nabla\alpha_2$ in $U$. 
Using the parametrization \eqref{Cl}, system \eqref{QS} reduces to 
\begin{equation}
\lr{\nabla\beta_1\cp\nabla\beta_2}\cp\lr{\nabla u_1\cp\nabla u_2}=\nabla\zeta,~~~~\abs{\nabla\beta_1\cp\nabla\beta_2}^2=f_B\lr{u_1,u_2},~~~~\Psi=\Psi\lr{\beta_1,\beta_2}.\label{QS2}
\end{equation}
In going from \eqref{QS} to \eqref{QS2} we used the fact that the first and third equations in \eqref{QSa} are identically satisfied. Furthermore, assuming $\bol{u}\neq\bol{0}$, 
the third equation in \eqref{QSa} implies that the modulus $B^2$ must be a function $f_B\lr{u_1,u_2}$ of $u_1$ and $u_2$.
Similarly, assuming that the magnetic field $\bol{B}\neq\bol{0}$ lies on flux surfaces 
one has $\bol{B}\cdot\nabla\Psi=0$ in $\Omega$, which implies that $\Psi$ must be a function of $\beta_1$ and $\beta_2$. 
The condition $\Psi=\Psi\lr{\beta_1,\beta_2}$ also ensures 
that boundary conditions \eqref{QSb} are fulfilled because $\bol{n}=\nabla\Psi/\abs{\nabla\Psi}$. 

Now our task is to solve system \eqref{QS2} by determining $\beta_1$, $\beta_2$, $u_1$, $u_2$, $f_B$, $\zeta$, and $\Psi$  
so that the level sets of $\Psi$ define toroidal surfaces. 
Direct integration of \eqref{QS2} is a mathematically difficult task due to the 
number and complexity of the geometric constraints involved. Therefore, it is convenient to start from known special solutions corresponding to axially symmetric configurations, and then perform a tailored symmetry breaking generalization. 
The simplest axially symmetric vacuum magnetic field is given by
\begin{equation}
\bol{B}_0=\nabla\varphi=\nabla z\cp\nabla\log r.\label{B0}
\end{equation}
The magnetic field \eqref{B0} satisfies system \eqref{QS} if, for example, the quasisymmetry is chosen as $\bol{u}_0=\bol{B}_0$. The corresponding flux surfaces are given by axially symmetric tori generated by level sets of the function 
\begin{equation}
\Psi_0=\frac{1}{2}\left[\lr{r-r_0}^2+z^2\right],\label{Psi0}
\end{equation}
with $r_0$ a positive real constant representing the radial position of the toroidal axis (major radius). 
Comparing equation \eqref{Cl} with equations \eqref{B0} and \eqref{Psi0}, 
one sees that $\beta_1=u_1=z$, $\beta_2=u_2=\log r$, $B_0^2=1/r^2=e^{-2u_2}$, and $\Psi_0=\frac{1}{2}\left[\lr{e^{\beta_2}-r_0}^2+\beta_1^2\right]$.  

\begin{figure}[h]
\hspace*{-0cm}\centering
    \includegraphics[scale=0.55]{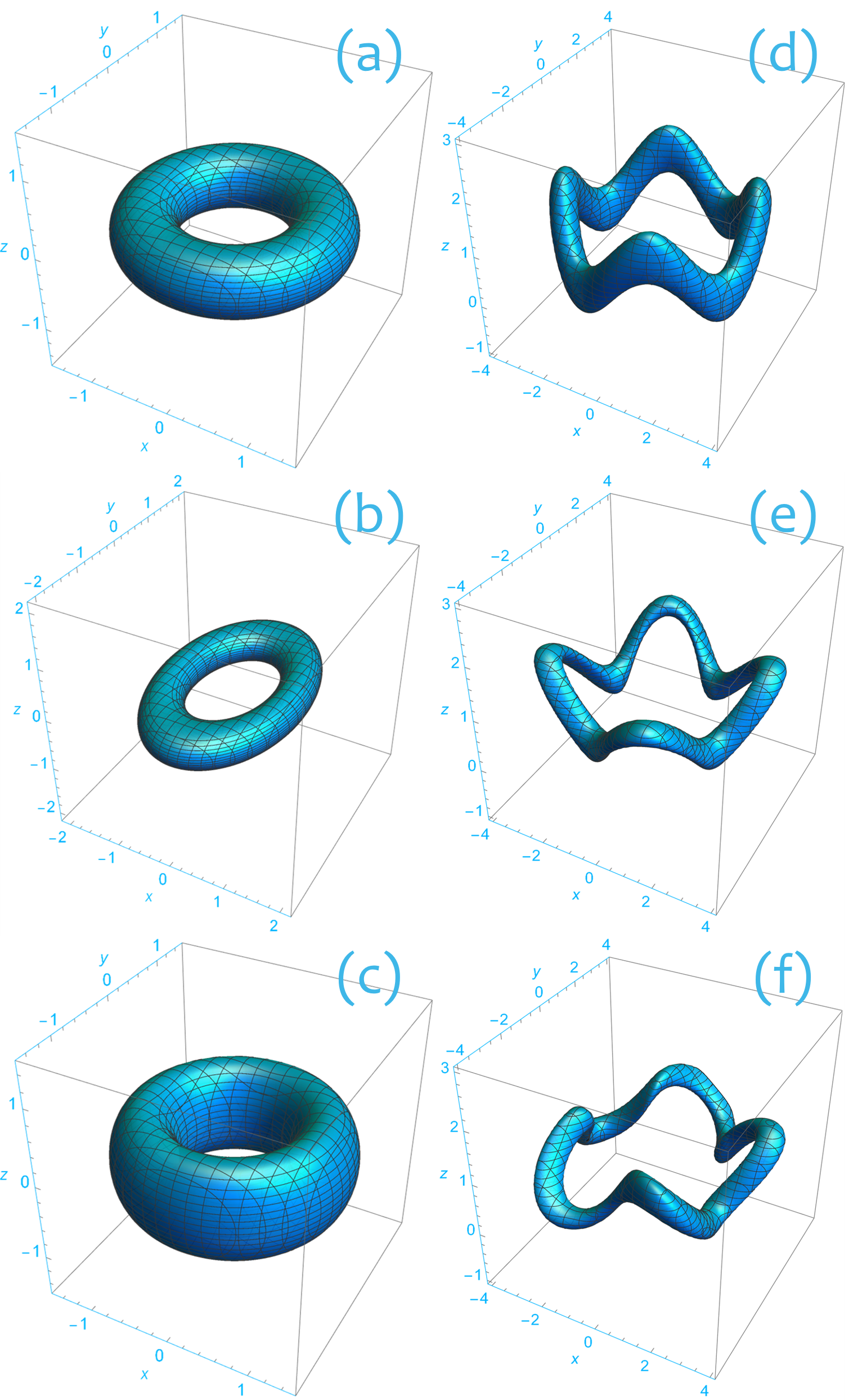}
\caption{\footnotesize Toroidal surfaces obtained as level sets of the function $\Psi$ defined by equation  \eqref{Psi}. (a) Axially symmetric torus $\Psi=0.15$ with $\mu=r$, $\mu_0=1$, $\mc{E}=1$, and $h=0$. (b) Elliptic torus $\Psi=0.1$ with $\mu=\sqrt{x^2+0.4 y^2}$, $\mu_0=1$, $\mc{E}=1$, and $h=0$. Notice that sections $z={\rm constant}$ form ellipses. (c) Axially symmetric torus $\Psi=0.15$ with $\mu=r$, $\mu_0=1$, $\mc{E}=0.4$, and $h=0$. Notice that sections $\varphi={\rm constant}$ form ellipses. (d) Torus $\Psi=0.1$ with $\mu=r$, $\mu_0=3$, $\mc{E}=1$, and $h=1+0.5\sin\lr{4\varphi}$. (e) Torus $\Psi=0.1$ with $\mu=r$, $\mu_0=3+0.5\sin\lr{4\varphi}$, $\mc{E}=5+2.5\sin\lr{4\varphi}$, and $h=1+0.5\sin\lr{4\varphi}$. (f) Torus $\Psi=0.1$ with $\mu=\sqrt{x^2+(0.9+0.1\sin\lr{3\varphi})y^2}$, $\mu_0=3+0.5\sin\lr{5\varphi}$, $\mc{E}=5+2.5\cos\lr{3\varphi}$, and $h=1+0.5\sin\lr{4\varphi}$.}
\label{fig2}
\end{figure}

The axially symmetric torus \eqref{Psi0} can be generalized to a larger class of toroidal surfaces \cite{Sato21} as 
\begin{equation}
\Psi=\frac{1}{2}\left[\lr{\mu-\mu_0}^2+\mc{E}\lr{z-h}^2\right]. \label{Psi}
\end{equation}
In this notation, $\mu$, $\mu_0$, $\mc{E}$, and $h$ are single valued functions with the following properties.
For each $z$, the function $\mu$ measures the distance of a point in the $\lr{x,y}$ plane from the origin in $\mathbb{R}^2$. 
The simplest of such measures is the radial coordinate $r$. More generally, on each plane $z={\rm constant}$ level sets of $\mu$ may depart from circles
and exhibit, for example, elliptical shape.  
The function $\mu_0$ assigns the $\mu$ value at which the toroidal axis is located. 
For the axially symmetric torus $\Psi_0$, we have $\mu_0=r_0$.  
The function $\mc{E}>0$ expresses the departure of toroidal cross sections (intersections of the torus with level sets of the toroidal angle) from circles. 
For example, the axially symmetric torus $\Psi_{\rm ell}=\frac{1}{2}\left[\lr{r-r_0}^2+2z^2\right]$ corresponding to $\mc{E}=2$ 
has elliptic cross section. Finally, the function $h$ can be interpreted as a measure of the vertical displacement of the toroidal axis from the $\lr{x,y}$ plane. Figure \ref{fig2} shows different toroidal surfaces generated through \eqref{Psi}.

The axial symmetry of the torus $\Psi_0$ given by \eqref{Psi0} can be broken by introducing dependence on the toroidal angle $\varphi$ in one of the functions $\mu$, $\mu_0$, $\mc{E}$, or $h$ appearing in \eqref{Psi}. 
Let us set $\mu=r$, take $\mu_0$ and $\mc{E}$ as positive constants, and consider 
a symmetry breaking vertical axial displacement $h=h\lr{r,\varphi,z}$.   
For the corresponding $\Psi$ to define a toroidal surface, the function $h$ must be single valued.
Hence, $\varphi$ must appear in $h$ as the argument of a periodic function. 
The simplest ansatz for $h$ is therefore
\begin{equation}
h=\epsilon \sin\left[m\varphi+g\lr{r,z}\right].\label{h}
\end{equation}
Here $m\in\mathbb{Z}$ is an integer, $\epsilon$ a positive control parameter such that 
the standard axially symmetric magnetic field $\bol{B}_0$ with flux surfaces $\Psi_0$ can be recovered in the limit $\epsilon\rightarrow 0$, and $g$ a function of $r$ and $z$ to be determined.  
Now recall that from equation \eqref{QS2} the function $\Psi$ is related to the Clebsch potentials $\beta_1$ and $\beta_2$ generating the magnetic field $\bol{B}=\nabla\beta_1\cp\nabla\beta_2$ according to   $\Psi\lr{\beta_1,\beta_2}$. Comparing with the axially symmetric case \eqref{Psi0} 
we therefore deduce that the analogy holds if $\beta_1=z-h$ and $\beta_2=\log r$. 
Defining $\eta=m\varphi+g$, it follows that the candidate quasisymmetric magnetic field is
\begin{equation}
\bol{B}=\nabla\left(z-\epsilon\sin\eta\right)\cp\nabla\log r=\left(1-\epsilon \cos\eta\frac{\p g}{\p z}\right)\nabla\varphi+\epsilon m\frac{\cos\eta}{r^2}\nabla z,\label{BQS1}
\end{equation}
where $g$ must be determined by enforcing quasisymmetry.
Next, observe that
\begin{equation}
B^2=\frac{1}{r^2}\left[\epsilon^2m^2\frac{\cos^2\eta}{r^2}+\lr{1-\epsilon\cos\eta\frac{\p g}{\p z}}^2\right].\label{B2}
\end{equation}
An essential feature of quasisymmetry \eqref{QS2} is that the modulus $B^2$ can be written 
as a function of two variables only, $B^2=f_B\lr{u_1,u_2}$. 
From equation \eqref{B2} one sees that this result can be achieved by setting $\p g/\p z=q\lr{r}$ 
for some radial function $q\lr{r}$ so that $u_1=\eta$, $u_2=\log r$, and also
\begin{equation}
g\lr{r,z}=q\lr{r} z+v\lr{r},
\end{equation}
with $v\lr{r}$ a radial function. 
The candidate direction of quasisymmetry is therefore
\begin{equation}
\bol{u}=\sigma\lr{\eta,r}\nabla\eta\cp\nabla\log r=\sigma\lr{\eta,r}\lr{q\nabla\varphi-\frac{m}{r^2}\nabla z},\label{uQS1}
\end{equation}
with $\sigma\lr{\eta,r}$ a function of $\eta$ and $r$ to be determined. 
Since by construction $B^2=B^2\lr{u_1,u_2}$, $\Psi=\Psi\lr{\beta_1,\beta_2}$, and both $\bol{B}$ and $\bol{u}$ as given by \eqref{BQS1} and \eqref{uQS1} are solenoidal, the only remaining equation in system \eqref{QS2} to be satisfied is the first one. In particular, we have
\begin{equation}
\bol{B}\cp\bol{u}=\sigma\lr{\nabla\varphi-\epsilon\cos\eta\nabla\eta\cp\nabla\log r}\cp\lr{\nabla\eta\cp\nabla\log r}=-m\frac{\sigma}{r^3}\nabla r.  
\end{equation}
Hence, upon setting $\sigma=\sigma\lr{r}$, system \eqref{QS2} is satisfied with 
\begin{equation}
\zeta=-m\int\frac{\sigma}{r^3}dr.
\end{equation}
Without loss of generality, we may set $\sigma=-r^3$ so that $\zeta=mr$ 
and the quasisymmetric configuration is given by
\begin{subequations}
\begin{align}
\bol{B}=&\nabla\left[z-\epsilon\sin\lr{m\varphi+qz+v}\right]\cp\nabla\log r=\left[1-\epsilon\cos\lr{m\varphi+qz+v}q\right]\nabla\varphi+\epsilon m\frac{\cos\lr{m\varphi+qz+v}}{r^2}\nabla z,\label{QS3a}\\
\bol{u}=&-\frac{1}{3}\nabla\lr{m\varphi+qz}\cp\nabla r^3=m r\nabla z-qr^3\nabla\varphi,\label{QS3b}\\
\Psi=&\frac{1}{2}\left\{\lr{r-r_0}^2+\mc{E}\left[z-\epsilon\sin\lr{m\varphi+qz+v}\right]^2\right\},\label{QS3c}
\end{align}\label{QS3}
\end{subequations}
where $\mc{E}$ is a positive real constant.

\section{Verification of asymmetry}
For the family of solutions \eqref{QS3} to qualify both as quasisymmetric 
and without continuous Euclidean isometries, we must verify that the magnetic field
\eqref{QS3a} is not invariant under some appropriate combination of translations and rotations. 
To see this, consider the case $q=1/r$ and $v=0$ corresponding to
\begin{subequations}
\begin{align}
\bol{B}=&\nabla\left[z-\epsilon\sin\lr{m\varphi+\frac{z}{r}}\right]\cp\nabla\log r=\left[1-\epsilon\frac{\cos\lr{m\varphi+\frac{z}{r}}}{r}\right]\nabla\varphi+\epsilon m\frac{\cos\lr{m\varphi+\frac{z}{r}}}{r^2}\nabla z,\label{QS4a}\\
\bol{u}=&-\frac{1}{3}\nabla\lr{m\varphi+\frac{z}{r}}\cp\nabla{r^3}=m r\nabla z-r^2\nabla\varphi,\label{QS4b}\\
\Psi=&\frac{1}{2}\left\{\lr{r-r_0}^2+\mc{E}\left[z-\epsilon\sin\lr{m\varphi+\frac{z}{r}}\right]^2\right\}\label{QS4c},
\end{align}\label{QS4}
\end{subequations}
where $\mc{E}$ is a positive real constant.
Notice that the magnetic field \eqref{QS4a} is smooth in any  domain $V\subset\mathbb{R}^3$ not containing the vertical axis $r=0$. 
To exclude the existence of any continuous Euclidean isometry for \eqref{QS4a} it is sufficient to show that
the equation
\begin{equation}
\mf{L}_{\bol{\xi}}B^2=\bol{\xi}\cdot\nabla B^2=0,~~~~\bol{\xi}=\bol{a}+\bol{b}\cp\bol{x},\label{Esym}
\end{equation}
does not have solution for any choice of constant vector fields $\bol{a},\bol{b}\in\mathbb{R}^3$ with $\bol{a}^2+\bol{b}^2\neq 0$. Indeed, since $\bol{\xi}=\bol{a}+\bol{b}\cp\bol{x}$ 
represents the generator of continous Euclidean isometries, the impossibility of satisfying 
\eqref{Esym} prevents the magnetic field $\bol{B}$ from possessing translational, axial, or helical symmetry.
For further details on this point, see \cite{Sato21}. 
Next, introducing again $\eta=m\varphi+z/r$, from equation \eqref{QS4a} one has
\begin{equation}
B^2=\frac{1}{r^2}-2\epsilon\frac{\cos\eta}{r^3}+\epsilon^2\lr{1+m^2}\frac{\cos^2\eta}{r^4}.\label{B22}
\end{equation}
It follows that
\begin{equation}
\begin{split}
\bol{\xi}\cdot\nabla B^2=&\frac{2}{r^3}\left[-1+3\epsilon\frac{\cos\eta}{r}-2\epsilon^2\lr{1+m^2}\frac{\cos^2\eta}{r^2}
\right]\bol{\xi}\cdot\nabla r\\&+2\epsilon\frac{\sin\eta}{r^3}\left[1-\epsilon\lr{1+m^2}\frac{\cos\eta}{r}\right]\bol{\xi}\cdot\nabla\eta.
\end{split}
\end{equation}
Let $\lr{a_x,a_y,a_z}$ and $\lr{b_x,b_y,b_z}$ denote the Cartesian components of $\bol{a}$ and $\bol{b}$.
On the surface $\eta=0$, corresponding to $z=z\lr{x,y}=-mr\varphi=-m\arctan\lr{y/x}\sqrt{x^2+y^2}$, we have $\sin\eta=0$ and $\cos\eta=1$, and therefore, 
\begin{equation}
\bol{\xi}\cdot\nabla B^2=\frac{2}{r^4}\left[-1+\frac{3\epsilon}{r}-2\epsilon^2\frac{\lr{1+m^2}}{r^2}\right]\left[xa_x+ya_y+\lr{xb_y-yb_x}z\lr{x,y}\right].
\end{equation}
This quantity vanishes provided that $a_x=a_y=b_x=b_y=0$. 
Consider now the surface $\eta=\pi/2$, which implies $z=z\lr{x,y}=r\lr{\pi/2-m\varphi}=\sqrt{x^2+y^2}\lr{\pi/2-m\arctan\lr{y/x}}$. In this case $\sin\eta=1$ while $\cos\eta=0$. 
Furthermore, since the only surviving components in $\bol{\xi}$ are those coming from $a_z$ and $b_z$, 
one has $\bol{\xi}\cdot\nabla r=0$, and therefore
\begin{equation}
\bol{\xi}\cdot\nabla B^2=\frac{2\epsilon}{r^3}\lr{\frac{a_z}{r}+mb_z}.
\end{equation}
This quantity vanishes provided that $a_z=b_z=0$.
Hence, the quasisymmetric magnetic field \eqref{QS4a} cannot possess continuous Euclidean isometries.  

Similarly, the flux function $\Psi$ defined by equation \eqref{QS4c} is not invariant under continuous Euclidean isometries. 
Indeed, the equation
\begin{equation}
\mf{L}_{\bol{\xi}}\Psi=\bol{\xi}\cdot\nabla\Psi=0,~~~~\bol{\xi}=\bol{a}+\bol{b}\cp\bol{x},
\end{equation}
does not have solution for any nontrivial choice of $\bol{a},\bol{b}\in\mathbb{R}^3$. 
This can be verified easily for $\abs{m}>1$. Indeed, in this case it is sufficient to evaluate $\bol{\xi}\cdot\nabla\Psi$ over the line $r=r_0$, $z=0$ parametrized by $\varphi$.
Here, we have
\begin{equation}
\begin{split}
\bol{\xi}\cdot\nabla\Psi=&-\epsilon\mc{E}\sin\lr{m\varphi}\bol{\xi}\cdot\nabla\lr{z-\epsilon\sin\eta}\\
=&-\epsilon\mc{E}\sin\lr{m\varphi}\left[ 
a_z-\frac{\epsilon a_z}{r_0}\cos\lr{m\varphi}+r_0b_x\sin\varphi-r_0b_y\cos\varphi-\epsilon\lr{b_x-\frac{ma_x}{r_0}}\sin\varphi\cos\lr{m\varphi}\right.
\\&\left.+\epsilon\lr{b_y-\frac{ma_y}{r_0}}\cos\varphi\cos\lr{m\varphi}-\epsilon m b_z\cos\lr{m\varphi}\right].
\end{split}
\end{equation}
This quantity identically vanishes provided that $a_x=a_y=a_z=b_x=b_y=b_z=0$. 

\section{Properties of the constructed solutions}
\begin{figure}[h!]
\hspace*{-0cm}\centering
    \includegraphics[scale=0.6]{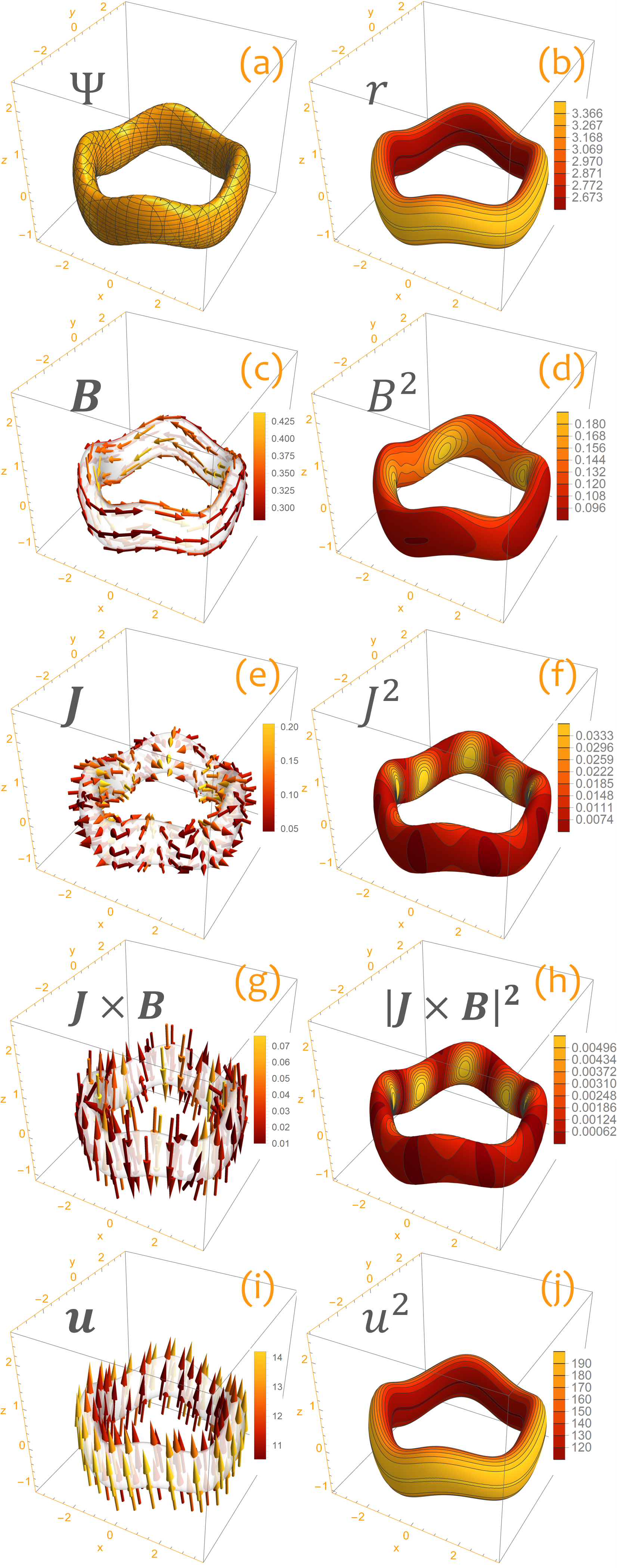}
\caption{\footnotesize The quasisymmetric configuration \eqref{QS4} for $r_0=3$, $\epsilon=0.2$, $m=4$ and $\mc{E}=0.7$. (a) Flux surface $\Psi=0.1$. (b) Levels sets of $r$ on the flux surface $\Psi=0.1$. These contours correspond to magnetic field lines. (c),  (d), (e), (f), (g), (h), (i), (j): plots of the magnetic field $\bol{B}$, the modulus $B^2$, the electric current $\bol{J}$, the modulus $J^2$, the Lorentz force $\bol{J}\cp\bol{B}$, the modulus $\abs{\bol{J}\cp\bol{B}}^2$, the quasisymmetry $\bol{u}$, and the modulus $u^2$ on the flux surface $\Psi=0.1$. }
\label{fig3}
\end{figure}

Let us examine the properties of the quasisymmetric configuration \eqref{QS4}.  
First, observe that level sets of \eqref{QS4c} define toroidal surfaces (see figure \ref{fig3}(a)), implying that the magnetic field \eqref{QS4a} has nested flux surfaces.
Next, note that the function $\zeta$ such that $\bol{B}\cp\bol{u}=\nabla \zeta$ is proportional to the radial coordinate, i.e. $\zeta=mr$. This function is associated with the conserved momentum $\bar{p}$ 
generated by the quasisymmetry. In particular, we have \cite{Rodriguez2020}
\begin{equation}
\bar{p}=-\frac{1}{\epsilon_{\rm gc}}\zeta+v_{\parallel}\frac{\bol{u}\cdot\bol{B}}{B}.\label{p}
\end{equation}
Here, $v_{\parallel}$ denotes the component of the velocity of a charged particle 
along the magnetic field $\bol{B}$ while $\epsilon_{\rm gc}\sim\rho/L$ is a small parameter associated with guiding center ordering, $\rho$ the gyroradius, and $L$ a characteristic length scale for the magnetic field.
It follows that charged particles moving in the magnetic field \eqref{QS4a} will approximately preserve their radial position since $\bar{p}\approx -\frac{m}{\epsilon_{\rm gc}}r$. This property works in favor of good confinement, although it cannot prevent particles from drifting in the vertical direction. The situation is thus analogous to the case of an axially symmetric
vacuum magnetic field $\bol{B}_{0}=\nabla\varphi$. 
Level sets of $\zeta=mr$ on a flux surface \eqref{QS4c} are shown in figure \ref{fig3}(b). 
These contours correspond to magnetic field lines because the magnetic field \eqref{QS4a} is such that $\bol{B}\cdot\nabla\Psi=\bol{B}\cdot\nabla r=0$, and field lines are solutions of the ordinary differential equation $\dot{\bol{x}}=\bol{B}$. In particular, observe that magnetic field lines are not twisted, and are given by the intersections of the surfaces $\Psi={\rm constant}$ and $r={\rm constant}$, implying that their projection on the $\lr{x,y}$ plane is a circle. 
Plots of the magnetic field \eqref{QS4a} and its modulus $B^2$ are given in figures \ref{fig3}(c) and \ref{fig3}(d). 
It is also worth noticing that the magnetic field \eqref{QS4a} is not a vacuum field. Indeed, it
has a non-vanishing current $\bol{J}=\nabla\cp\bol{B}$ given by
\begin{equation}
\bol{J}=\frac{\epsilon}{r^3}\left[-\lr{1+m^2}\sin\eta\nabla r+m\lr{2r\cos\eta-z\sin\eta}\nabla\varphi+\lr{\cos\eta-\frac{z}{r}\sin\eta}\nabla z\right].
\end{equation}
Figures \ref{fig3}(e) and \ref{fig3}(f) show plots of the current field $\bol{J}$ and the corresponding modulus $J^2$. 
The Lorentz force $\bol{J}\cp\bol{B}$ can be evaluated to be
\begin{equation}
\begin{split}
\bol{J}\cp\bol{B}=&\frac{\epsilon}{r^4}\left\{
\left[\lr{\cos\eta-\frac{z}{r}\sin\eta}\lr{\epsilon\lr{1+m^2}\frac{\cos\eta}{r}-1}+\epsilon m^2\frac{\cos^2\eta}{r}\right]\nabla r\right.\\&\left.+\epsilon m\lr{1+m^2}\sin\eta\cos\eta\nabla\varphi-\lr{1+m^2}\sin\eta\lr{1-\epsilon\frac{\cos\eta}{r}}\nabla z
\right\}.
\end{split}
\end{equation}
It is not difficult to verify that the right-hand side of this equation cannot be written as the gradient of a pressure field $\nabla P$. Hence, the quasisymmetric magnetic field \eqref{QS4a} does not represent an equilibrium of ideal magnetohydrodynamics. Nevertheless, it can be regarded as an equilibrium of anistropic magnetohydrodynamics provided that the 
components of the pressure tensor are appropriately chosen (on this point, see \cite{Sato21}). 
Plots of the Lorentz force $\bol{J}\cp\bol{B}$ and its modulus $\abs{\bol{J}\cp\bol{B}}^2$ are given in figures \ref{fig3}(g) and \ref{fig3}(h). 
Next, observe that the quasisymmetry $\bol{u}$ given by equation \eqref{QS4b} is not tangential to flux surfaces $\Psi$.
Indeed, 
\begin{equation}
\bol{u}\cdot\nabla\Psi=m\mc{E}\lr{z-\epsilon  \sin\eta}r. 
\end{equation}
Plots of the quasisymmetry $\bol{u}$ and its modulus $u^2$ can be found in figures \ref{fig3}(i) and \ref{fig3}(j). 

\begin{figure}[h!]
\hspace*{-0cm}\centering
    \includegraphics[scale=0.32]{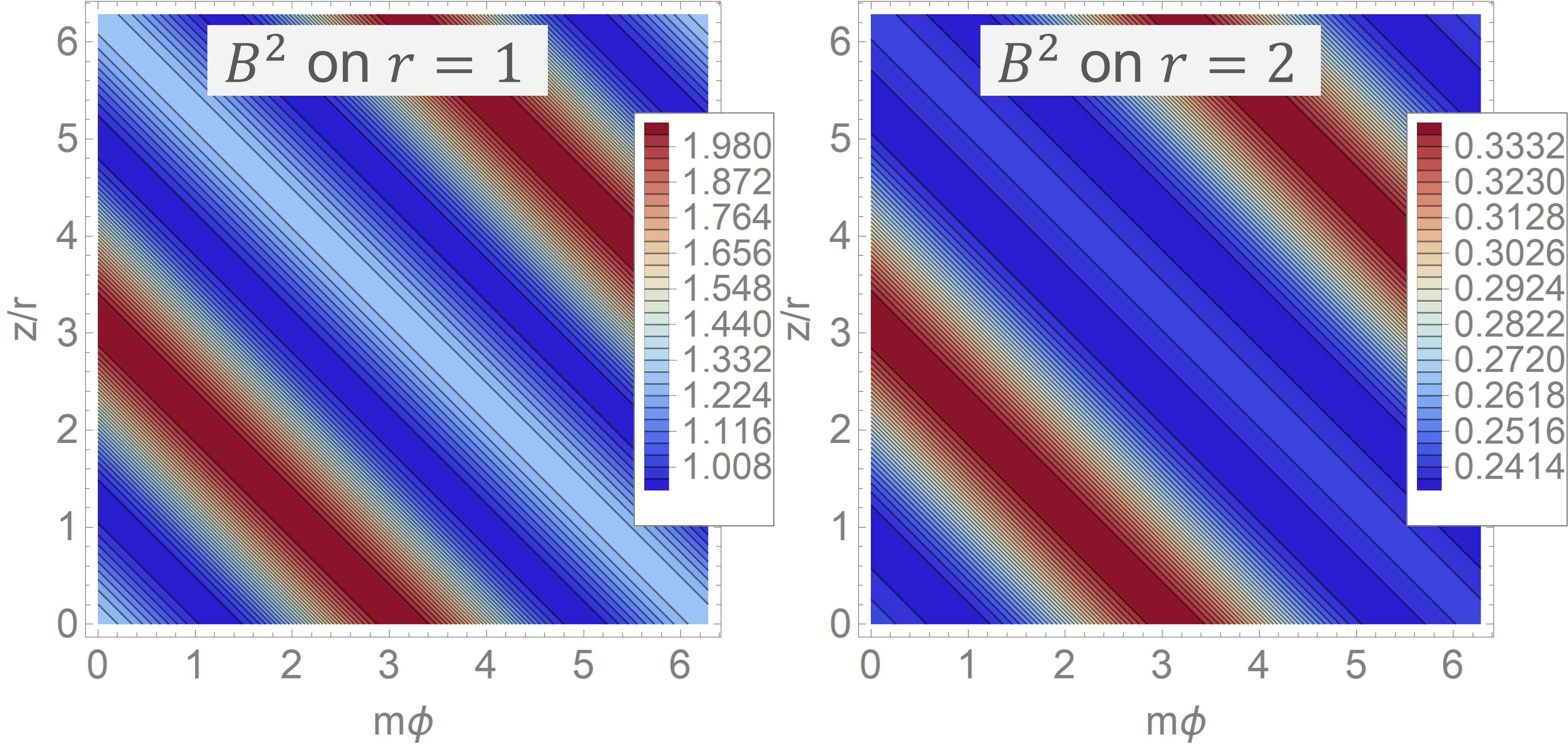}
\caption{\footnotesize Modulus $B^2\lr{r,m\varphi+z/r}$ of the quasisymmetric magnetic field \eqref{QS4a} for $\epsilon=0.2$ and $m=4$ as seen in the $\lr{m\varphi,z/r}$ plane for different values of the radial coordinate $r$. (a) Plot on the level set $r=1$. (b) Plot on the level set $r=2$. Observe how  contours of $B^2$ form straight lines.}
\label{fig4}
\end{figure}

\begin{figure}[h!]
\hspace*{-0cm}\centering
    \includegraphics[scale=0.32]{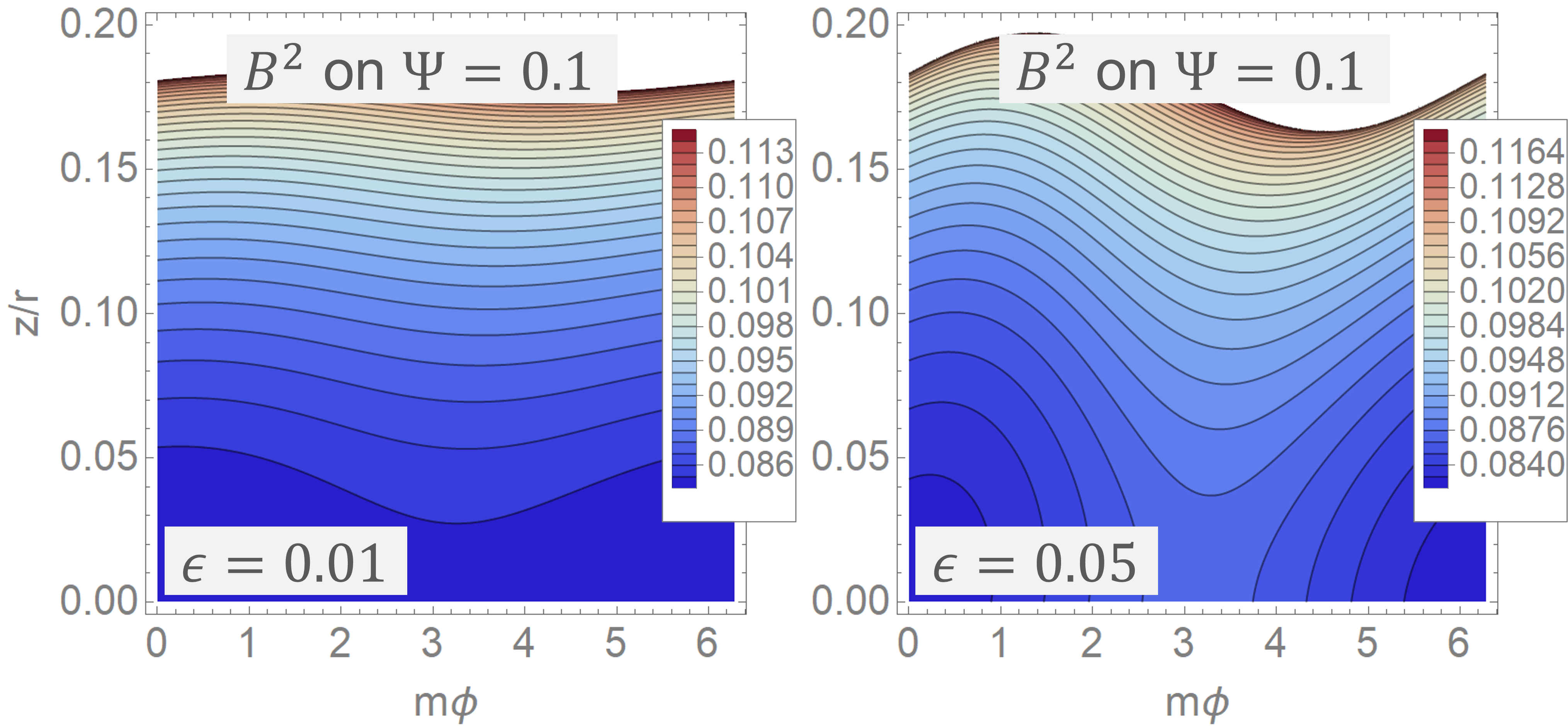}
\caption{\footnotesize Modulus $B^2\lr{r\lr{\Psi,z/r,\eta},\eta}$ with $\eta=m\varphi+z/r$ of the quasisymmetric magnetic field \eqref{QS4a} for $r_0=3$, $m=4$ and $\mc{E}=0.7$ as seen in the $\lr{m\varphi,z/r}$ plane corresponding to $\Psi=0.1$. (a) The case $\epsilon=0.01$. (b) The case $\epsilon=0.05$. Notice that white regions in the plot reflect the fact that for given values of $\Psi$ and $\varphi$ the range of $z$ is bounded.}
\label{fig5}
\end{figure}

Finally, let us consider how the quasisymmetry of the configuration \eqref{QS4} 
compares with the usual understanding that the modulus of a quasisymmetric magnetic field 
depends on the flux function $\Psi$ 
and a linear combination of toroidal angle $\varphi$ and poloidal angle $\vartheta$, 
i.e. $B^2\lr{\Psi,M\vartheta-N\varphi}$ with $M,N$ integers.
When $B^2=B^2\lr{\Psi,M\vartheta-N\varphi}$, on each flux surface 
the contours of the modulus $B^2$ in the $\lr{\varphi,\theta}$ plane form straight lines. 
For the quasisymmetric magnetic field \eqref{QS4a} we have $B^2=B^2\lr{r,m\varphi+z/r}$.
Hence, the correspondence 
with the usual setting can be obtained by the identification  $\Psi\rightarrow r$, $\varphi\rightarrow\varphi$, and $\vartheta\rightarrow z/r$.
Figure \ref{fig4} shows how the contours of the quasisymmetric magnetic field \eqref{QS4a} form straight lines in the $\lr{m\varphi,z/r}$ plane.
Next, it is useful to determine how much the contours of $B^2$ depart from straight lines
on each flux surface. To this end, observe that equation \eqref{QS4c} can be inverted to obtain 
$r\lr{\Psi,z/r,\eta}$ with $\eta=m\varphi+z/r$ so that the modulus \eqref{B22} can be written in the form $B^2=B^2\lr{r\lr{\Psi,z/r,\eta},\eta}$. 
Figure \ref{fig5} shows contours of $B^2$ on the plane $\lr{m\varphi,z/r}$
for a fixed value of $\Psi$ and different choices of the parameter $\epsilon$ controlling the
degree of asymmetry of the solution. 
In particular, notice how the solution \eqref{QS4} approaches axial symmetry for smaller values of $\epsilon$.

\section{Concluding Remarks}
In conclusion, we have demonstrated the existence of weakly quasisymmetric magnetic fields in toroidal volumes  
by constructing explicit examples \eqref{QS3} through the method of Clebsch parametrization.   
The obtained configurations are solutions of system \eqref{QS} with the following properties. 
In the optimized toroidal domain $\Omega$, 
the magnetic field $\bol{B}$ is smooth and equipped with nested flux surfaces $\Psi$.
Both $\bol{B}$ and $\Psi$ do not exhibit continuous Euclidean isometries, i.e. invariance under an appropriate combination of translations and rotations. 
The quasisymmetry $\bol{u}$ is not tangential to toroidal flux surfaces $\Psi$, but lies on surfaces of constant
radius $r$. In particular, $\bol{B}\cp\bol{u}=m\nabla r$ with $m$ an integer while $B^2=B^2\lr{r,m\varphi+z/r}$ in the example \eqref{QS4}.  
The conserved momentum arising from the quasisymmetry is given by \eqref{p},
which is approximately the radial position of a charged particle. 
The magnetic field $\bol{B}$ is not a vacuum field since a current $\bol{J}=\nabla\cp\bol{B}\neq\bol{0}$ is present. The obtained quasisymmetric magnetic fields \eqref{QS3a} can be regarded
as solutions of anisotropic magnetohydrodynamics if the 
component of the pressure tensor are appropriately chosen \cite{Sato21}. 

In addition to providing mathematical proof of existence of solutions to system \eqref{QS} with
the properties described above, 
this work offers an alternative theoretical framework for the numerical and experimental efforts devoted to modern stellarator design, and possibly paves the way to the development of semi-analytical schemes aimed at the optimization of confining magnetic fields. 
The next goal of the present theory would be to further improve the obtained results by 
ascertaining the existence of vacuum solutions $\nabla\cp\bol{B}=\bol{0}$ of system \eqref{QS} 
such that the modulus of the magnetic field can be written as a function of the flux function and a linear combination of toroidal and poloidal angles, $B^2=B^2\lr{\Psi,M\vartheta-N\varphi}$, 
and in particular to establish the existence of vacuum quasisymmetric configurations 
with the field line twist required to effectively trap charged particles.

\section*{Declarations}
\subsection*{Acknowledgments}
The research of NS was partially supported by JSPS KAKENHI Grants No. 21K13851 and No. 22H00115. The author acknowledges usueful discussion with Z. Qu, D. Pfefferl\'e, R. L. Dewar, T. Yokoyama, and with several members of the Simons Collaboration on Hidden Symmetries and Fusion Energy.

\subsection*{Author contributions}
N.S. developed the theoretical formalism, performed the analytic calculations, and wrote the manuscript.

\subsection*{Data availability} No datasets were generated or analyzed in this study. 

\subsection*{Competing interests} 
The author has no relevant financial or non-financial interests to disclose.

\end{document}